\documentclass{article}
\usepackage{amssymb,amsmath,amsthm,stmaryrd,mathrsfs}

\newfont{\Fr}{eufm10}
\newfont{\Sc}{eusm10}

\newcommand{\bb}{\mathbb}
\newcommand{\ms}{\mathscr}
\newcommand{\Rn}{{\bb R}^n}
\newtheorem{defn}{Definition}[section]
\newtheorem{Lem}[defn]{Lemma}
\newtheorem{Thm}[defn]{Theorem}

\newtheorem{Rem}[defn]{Remark}
\newtheorem{Prop}[defn]{Proposition}
\newtheorem{Cor}[defn]{Corollary}
\newtheorem{conjecture}[defn]{Conjecture}

\newtheorem{main theorem}[defn]{Main Theorem}

\input pictex

\usepackage{hyperref}
\begin{document}

\title{Legendre Transform, Hessian Conjecture and Tree Formula}
\author{Guowu Meng\\ \small{\it Department of Mathematics}\\ \small{\it Hong Kong University of Science and Technology}\\
\small{\it Clear Water Bay, Kowloon, Hong Kong}\\
\small{Email: mameng@ust.hk}}
\date{January 31, 2005}
\maketitle

\begin{abstract}
Let $\varphi$ be a polynomial over $K$ (a field of characteristic
$0$) such that the Hessian of $\varphi$ is a nonzero constant. Let
$\bar\varphi$ be the formal Legendre Transform of $\varphi$. Then
$\bar\varphi$ is well-defined as a formal power series over $K$.
The Hessian Conjecture introduced here claims that $\bar\varphi$
is actually a polynomial. This conjecture is shown to be true when
$K=\bb{R}$ and the Hessian matrix of $\varphi$ is either positive
or negative definite somewhere. It is also shown to be equivalent
to the famous Jacobian Conjecture. Finally, a tree formula for
$\bar\varphi$ is derived; as a consequence, the tree inversion
formula of Gurja and Abyankar is obtained.
\end{abstract}

\section{Introduction}
The Jacobian conjecture is one of the famous open fundamental problems
in mathematics \cite{Sma98}, and is very often stated as
\begin{conjecture}[Jacobian Conjecture]  Let $f$: $\bb{C}^n\rightarrow \bb{C}^n$
be a polynomial map whose
Jacobian is a nonzero constant, then $f$ is invertible and the
inverse is also a polynomial.
\end{conjecture}
(In fact the field $\bb{C}$ can be replaced by any field of
characteristic zero. But the analogue for a field with
characteristic $p>0$ is false. See reference \cite{BCW82}.)

Originally called Keller's problem \cite{Kel39}, the Jacobian
Conjecture  has a few published faulty proofs \cite{Eng55,
Seg56-57, Seg56, Seg60}. Over a hundred papers have been
published, but the conjecture is still open even in dimension two.
Like many other famous conjectures, this conjecture is deceptively
simple!

Reference \cite{BCW82} gives an excellent review on the Jacobian
Conjecture up to 1982. For a more recent review and references on
the Jacobian Conjecture, the reader may consult reference
\cite{Ess01}.

\vskip 10pt It is probably well-known to people working on the
Jacobian Conjecture that there are many other conjectures which
are equivalent to the Jacobian Conjecture. Here we propose another
equivalent conjecture --- the Hessian Conjecture. This conjecture
grows out of the author's failed attempt to settle the Jacobian
Conjecture and is interesting in its own right; and it looks
simpler: instead of dealing with many polynomials, one just needs
to deal with a single polynomial.

\vskip 10pt Our thanks to  A. Voronov for introducing us to the
one-dimensional tree inversion formula and reference \cite{BCW82}.
A discussion of the tree inversion formula of Gurja and Abyankar
in terms of Feynman diagrams has recently appeared in
\cite{Abd02a,Abd02b}, but our discussion has a different
perspective and our  proof and derivation of the tree formula use
somewhat different ideas. Significant work has appeared pertaining
to polynomial maps with symmetric Jacobian matrices
\cite{JC4SymmetricJ,NilpotentSymmetric,SymmetricReduction}. In
\cite{SymmetricReduction} M. de Bondt and A. van den Essen
describe a Hessian conjecture virtually identical to the one we
formulate, but which does not involve the Legendre transform. They
also show its equivalence to the Jacobian conjecture and prove the
reduction theorem of section \ref{reduc-sec} in this paper. We
learned of that work only after this paper was originally written.
Although the Hessian conjecture has been articulated as such only
recently, the first result in this area - the case $n=2$ - was
proved in 1991 \cite{SymmetricJacobian}. This work is supported by
the Hong Kong Research Grants Council under the RGC project
HKUST6161/97P.

\subsection{ Hessian Conjecture}
Let $K$ be a field of characteristic zero, $\varphi$ a polynomial
in $n$ variables with coefficients in $K$, i.e., $\varphi\in
K[x_1,\cdots, x_n]$. The Hessian matrix $H_\varphi(x)$ is a
symmetric matrix whose $(i,j)$-entry is
$\partial_i\partial_j\varphi(x)$. By definition, the determinant
of $H_\varphi(x)$ is called the Hessian of $\varphi$ at $x$,
denoted by $h_\varphi(x)$.

Suppose that $h_\varphi\neq 0$ at $x=0$, then
$y=\nabla\varphi(x):=(\partial_1
\varphi(x),\cdots,\partial_n\varphi(x))$ has a formal inverse
$x=g(y)$ --- a formal power series in $y$. Let $\bar \varphi(y)$
be the (formal) Legendre transform of $\varphi$, i.e., $\bar
\varphi(y)$ is a formal power series in $y$ defined by
equation
\begin{eqnarray}\label{LT}
\bar\varphi(y)=[xy-\varphi(x)]|_{x=g(y)}.
\end{eqnarray}
It is clear that $x=\nabla\bar\varphi(y)$, so $\bar\varphi$ is a
potential function for $g$. Obviously $\bar\varphi$ is a formal
power series in $y$; however, we may consider the

\begin{conjecture}[Hessian Conjecture]\label{HC} Let $\varphi$ be a polynomial
over $K$ whose
Hessian is a nonzero constant, $\bar\varphi$ the formal Legendre
transform of $\varphi$. Then $\bar\varphi$ is also a polynomial.
\end{conjecture}

\begin{Thm} The Hessian Conjecture is true when $K=\bb{R}$ and the Hessian matrix
is definite (either positive or negative) somewhere. Therefore, if
\[
\varphi(x)={1\over 2} x^2+\mbox{higher order terms}
\]
is a real polynomial with $h_{\varphi}=1$ everywhere, then
$\bar\varphi$ is also a polynomial.
\end{Thm}
\begin{proof} Let $\varphi$ be a real polynomial function on
$\bb{R}^n$ whose Hessian is conatant. Without the loss of
generality we may assume $h_\varphi=1$ everywhere.

\smallskip
{\bf Claim 1}: $H_\varphi$ is non-degenerate everywhere and has
constant signature. Therefore, if $H_\varphi$ is positive
(negative) definite somewhere, it is positive (negative) definite
everywhere.

\noindent {\bf Proof of the claim 1}. Fix $x\in \bb{R}^n$. Define
\[
O(t):=H_\varphi(tx).
\]
Then $O$ is a smooth path in the space of nondegenerate (because
of the Hessian condition on $\varphi$),  real symmetric $n\times
n$ matrices; therefore we have a spectral flow from $t=0$ to
$t=1$. The Hessian condition on $\varphi$ implies that the
signature of $O(1)=H_\varphi(x)$ must be equal to that of $O(0)$;
otherwise, there would be a zero eigenvalue somewhere along the
path, say at $t_0$ ($0<t_0<1$), but then we would have the
following contradiction:
\begin{eqnarray}
0=\det O(t_0)=\det H_\varphi(t_0x).\nonumber
\end{eqnarray}
\smallskip
{\bf Claim 2}. As a map from $\bb{R}^n$ to $\bb{R}^n$,
$\nabla\varphi$ is one to one.

\noindent {\bf Proof of claim 2}. Suppose that
$\nabla\varphi(x_1)=\nabla\varphi(x_2)$ for some points $x_1$ and
$x_2$ in $\bb{R}^n$. Set $f(t)=(x_2-x_1)\cdot
\nabla\varphi(x_1+t(x_2-x_1))$ for $0\le t\le 1$. Note that
$f(0)=f(1)$, so there is a $t_0\in (0,1)$ such that $f'(t_0)=0$,
i.e., $$(x_2-x_1)^T H_\varphi(x_1+t_0(x_2-x_1)) (x_2-x_1)=0.$$ By
the assumption on $\varphi$ and claim 1 above, we know that
$H_\varphi(x_1+t_0(x_2-x_1))$ is definite, so $x_2-x_1=0$, i.e.,
$x_2=x_1$.
\medskip Since $\nabla\varphi$ is one to one, by Theorem 2.1 of
reference \cite{BCW82}, we know $\nabla\varphi$ has a polynomial
inverse, so it is clear from equation (\ref{LT}) that
$\bar\varphi$ is also a polynomial.
\end{proof}

\begin{Prop}\label{equiv}
The Hessian Conjecture is equivalent to the Jacobian Conjecture.
\end{Prop}
\begin{proof} If the Jacobian Conjecture is true, then equation
(\ref{LT}) implies that the Hessian Conjecture is also true. On
the other hand, assume the Hessian Conjecture is true, then the
Jacobian Conjecture is also true, and this can be proved by the
following trick: Let $f$: $K^n\rightarrow K^n$ be a polynomial map
whose Jacobian is $1$ everywhere. Let $\varphi (v,x)=v\cdot f(x)$,
then $\varphi$ is a polynomial function on $K^{2n}$ whose Hessian
is $(-1)^n$ everywhere. Then $\bar\varphi$ is also a polynomial
function by the assumption. Now $\bar\varphi(w,y)=w\cdot
f^{-1}(y)$ where $f^{-1}(y)$ is the formal inverse of $f$, so
$f^{-1}(y)$ is also a polynomial.
\end{proof}

\subsection{A Reduction Theorem}\label{reduc-sec}
In view of the reduction theorem in \cite{BCW82} and the proof of
Proposition (\ref{equiv}), the following reduction theorem can be
easily deduced.

\begin{Thm} The Hessian Conjecture is true $\Leftrightarrow$  for each integer $n\ge
1$ and for each polynomial map $\varphi$ : $\bb{C}^{2n}\to\bb{C} $
of the form
\[
\varphi(x)={1\over 2}x^2+\hbox{a homogeneous quartic polynomial in
$x$},
\]
if the hessian of $\varphi$ is constant, then $\bar\varphi$ is a
polynomial.
\end{Thm}
In section \ref{tree} we shall introduce and prove a tree formula
for $\bar\varphi$; as a consequence, we obtain the tree formula of
Gurja and Abyankar \cite{Abh74, BCW82}.

\section{A Tree Formula}\label{tree}
Let  $x=(x^1,..., x^n)$ and
\begin{eqnarray}\label{varphitype}
\varphi(x)=\sum_{2N\ge m\ge 2}{1\over m!}T_m( x),
\end{eqnarray}
where $N>1$ is an integer and $T_m(x)$ is a degree $m$ homogeneous
polynomial in $x$. Note that $T_m(x)$ should be identified with
$T_m=[(T_m)_{i_1\ldots i_m}]$---a symmetric tensor of $m$ indices:
$$T_m(x)=(T_m)_{i_1\ldots i_n}x^{i_1}\cdots x^{i_m}.$$
(Here the repeated indices are summed up.)

Assume that $T_2$ is non-degenerate. Then we can introduce
the symmetric tensor ${T_2}^{-1}$: by definition,
$[({T_2}^{-1})^{ij}]$ is the inverse matrix of $[(T_2)_{ij}]$.
Under the assumption, we can formally solve equation $
y=(\partial_1\varphi( x),\ldots,
\partial_n\varphi(x))$ for $x$, so the Legendre
transformation (\ref{LT}) is well-defined. We say $\varphi$ is
non-degenerate if its degree two homogeneous component is
non-degenerate.

\begin{Thm}[Tree Formula]\label{Formula} Suppose that $\varphi$ is non-degenerate,
then the formal Legendre transform of $\varphi$ has the following
tree expansion formula:
\begin{eqnarray}\label{treeformula}
\bar\varphi(y)=\sum_{\Gamma\in\left\{\mbox{connected tree
diagrams}\right\}}w(\Gamma)
\end{eqnarray}
where $w(\Gamma)$ is the contribution from tree diagram $\Gamma$
and is given according to the following rules:

1) to each edge of $\Gamma$, assign $T_2^{-1}$,

2) to each external vertex, we assign $y=(y_1,\cdots,y_n)$,

3) to each internal vertex of degree $n$ assign $-T_n$,

4) multiply all assignments  in 1) through 3) and make all
necessary contractions and  then divided by $|\mathrm{Aut}
\Gamma|$ to get $w(\Gamma)$.
\end{Thm}
Here $\mathrm{Aut}(\Gamma)$ is the automorphism group of $\Gamma$
(see the appendix for its precise meaning) and
$|\mathrm{Aut}(\Gamma)|$ is the order of $\mathrm{Aut}(\Gamma)$.

To help the readers to understand the rules in the theorem, let us
present two examples here:

{\bf Example 1}. \vskip 10pt \hskip 100pt
\beginpicture
\setcoordinatesystem units <7pt, 7pt> \setlinear \plot 0 0  0 10 /
\put{$y$} at 0 11 \put{$y$} at 0 -1 \put{$T_2^{-1}$} at 1.5 5
\put{$={1\over 2}y_iy_j(T_2^{-1})^{ij},$} at 9 5
\endpicture
\vskip 20pt {\bf Example 2}. \vskip 15pt \hskip 10pt
\beginpicture
\setcoordinatesystem units <5pt, 5pt> \setlinear \plot 0 0  0 -10
/ \setlinear \plot 0 0  8 5 / \setlinear \plot 0 0  -8 5 /
\put{$y$} at 9 6 \put{$y$} at -9 6 \put{$y$} at 0 -12 \put{$-T_3$}
at 2 -2 \put{$T_2^{-1}$} at 6 1 \put{$T_2^{-1}$} at -5 1
\put{$T_2^{-1}$} at -2 -6 \put{$={1\over
3!}y_{i_1}y_{i_2}y_{i_3}(-T_3)_{j_1j_2j_3}(T_2^{-1})^{i_1j_1}(T_2^{-1})^{i_2j_2}(T_2^{-1})^{i_3j_3}$,}
at 32 0
\endpicture
\vskip 20pt \noindent where the repeated indices are summed up.

\vskip 10pt Write $\bar\varphi(y)=\sum_{m\ge 2} {1\over m!}S_m(y)$
and the right-hand side of (\ref{treeformula}) as $\sum_{m\ge 2}
{1\over m!}\tilde S_m(y)$, where both $S_m(y)$ and $\tilde S_m(y)$
are degree $m$ homogeneous polynomials in $y$. It is not hard to
see that each coefficient $C$ of $S_m(y)-\tilde S_m(y)$ is a
rational function (over the field of rational numbers) in the
coefficients of $T_2$, ..., $T_{2N}$. To prove
(\ref{treeformula}), we need to show that each $C$ is zero as a
rational function in the coefficients of $T_2$, ..., $T_{2N}$,
equivalently, we need to show that the zero set of each $C$
contains an open subset. Therefore, without the loss of
generality, we may assume that $K=\bb{R}$; moreover, we just need
to show that each $C$ has value zero for all $\varphi$ in a
non-empty open set of ${\cal P}_N$---the space all real
polynomials without linear and constant terms and having degree at
most $2N$, i.e., we just need to prove Theorem \ref{Formula} for
all $\varphi$ in an non-empty open set of ${\cal P}_N$. ( ${\cal
P}_N$ is a vector space and we can put a metric on it: by
definition, if $f$, $g$ are in ${\cal P}_N$, then the distance
between $f$ and $g$ is defined to be the maximum of the absolute
value of the coefficients of $f-g$.)

\begin{Lem}\label{key}
There is an non-empty open set $U_N$ in ${\cal P}_N$ such that for
all $\varphi\in U_N$ we have 1) $\varphi(x)>{1\over 2}|x|^{2N}$ if
$|x|$ is sufficiently large; 2) $0$ is the only critical point of
$\varphi$; 3) the quadratic component $T_2$ of $\varphi$ is
positive definite.
\end{Lem}
\begin{proof} Let $\varphi_0(x)=(|x|^2+1)^N-1$. Then $\varphi_0$
satisfies conditions 1), 2) and 3) in the lemma. It is not hard to
see that if $\varphi$ is sufficiently close to $\varphi_0$, then
$\varphi$ satisfies conditions 1), 2) and 3) in the lemma, too. So
we can take $U_N$ to be a sufficiently small ball centered at
$\varphi_0$.
\end{proof}
\begin{Cor}
Theorem \ref{Formula} is valid for all $\varphi\in U_N$.
\end{Cor}
\begin{proof} Assume $\varphi\in U_N$. Without loss of generality, we may assume
$T_2(x)=|x|^2$ --- that amounts to a rotation of the coordinate
system.

The proof is obtained by evaluating
\begin{eqnarray}\label{exp}
\lim_{\hbar \to 0}\hbar \log {\int dx \exp{1\over
\hbar}{\left(yx-\varphi(x)\right)}\over \int dx
\exp{\left(-{1\over 2\hbar}T_2(x)\right)} }
\end{eqnarray}
in two different ways. (The integrations are done over the whole
space $\Rn$.)

On the one hand, assume $|y|$ is sufficiently small, using the
assumption on $\varphi$, by the steepest decent \cite{BH86}, this
limit becomes
\begin{eqnarray}
yz-\varphi(z),
\end{eqnarray}
where $z$ is the unique solution of equation
\begin{eqnarray}
y-\nabla\varphi(x)=0
\end{eqnarray}
for $x$, i.e., $z=(\nabla\varphi)^{-1}(y)$. Therefore, this limit
is $\varphi'(y)$---the Legendre Transform of $\varphi$ as a
function (not as a formal power series) in $y$ and the
coefficients of $T_3$, ..., $T_{2N}$.

On the other hand, using the assumption on $\varphi$, we can
calculate
\begin{eqnarray}
\hbar\log {\int dx \exp{1\over
\hbar}{\left(yx-\varphi(x)\right)}\over \int dx
\exp{\left(-{1\over 2\hbar}T_2(x)\right)} }
\end{eqnarray}
in terms of connected Feynman diagrams to get its asymptotic
series expansion\footnote{For a definition of asymptotic series
expansion, see reference \cite{BH86}.} in $\hbar$, $y$ and the
coefficients of $T_3$, ..., $T_{2N}$, see the appendix for more
details. Note that the contribution from a connected Feynman
diagram with $m$ loops is proportional to $\hbar^{m}$, so only the
contributions from the tree diagrams survive in limit (\ref{exp}).
Since the contributions from the tree diagrams are exactly given
by the rules specified in Theorem \ref{Formula}, we have the
right-hand side of (\ref{treeformula}) which can be seen to be an
asymptotic series expansion for $\varphi'$ in $y$ and the
coefficients of $T_3$, ..., $T_{2N}$.

By the definition of $\bar\varphi$ and $\varphi'$, one can see
that $\bar\varphi$ is a convergent power series expansion of
$\varphi'$, hence it is also an asymptotical series expansion for
$\varphi'$ in $y$ and the coefficients of $T_3$, ..., $T_{2N}$. By
the uniqueness of asymptotical series expansion, we have a proof
of Theorem \ref{Formula} for $\varphi\in U_N$.
\end{proof}

\noindent {\bf Proof of Theorem \ref{Formula}}. The proof in the
general case follows from the above corollary and the discussion
preceding to Lemma \ref{key}. \hfill $\Box$
\begin{Rem}
Strictly speaking, we should do some estimates to fully justify
some of the arguments in the above proof of Lemma \ref{key} and
its corollary. These estimates are not hard to obtain; however,
they would make the paper lengthy and also make the main ideas
behind the proof a little bit obscure.
\end{Rem}

\begin{Rem}
Using the trick involved in the proof of Proposition \ref{equiv},
it is not hard to see that the tree formula given in this paper
and the tree formula of Gurja and Abyankar actually imply each
other. While the original proof of the tree formula of Gurja and
Abyankar is purely algebraic, the proof given here for our tree
formula is both algebraic and analytic.
\end{Rem}

\appendix

\section{Feynman Diagrams for Lebesgue integrals}
A very good reference for the discussion below is \cite{BIZ80}.
Let
\begin{equation}
Y(\lambda)=\int dx\exp\left(-{a\over 2}x^2-{\lambda\over
4!}x^4\right)\equiv \left\langle \exp\left(-{\lambda\over
4!}x^4\right)\right\rangle
\end{equation}
where $a>0$ and $\lambda>0$ are parameters and the integration is
done  over $\bb R$ and the integration measure is normalized so
that
\begin{equation}
\int dx\exp\left(-{a\over 2}x^2\right)=1.
\end{equation}
We are interested in the perturbative computation of $Y(\lambda)$.
Formally, we have
\begin{equation}
Y(\lambda)\sim \sum_n {1\over n!(4!)^n}(-\lambda)^n\left\langle
\underbrace{x^4\cdots x^4}_n\right\rangle,
\end{equation}
where symbol $\sim$ means the asymptotic series expansion of
$Y(\lambda)$ as $\lambda\to 0$. We would like to compute
$${1\over n!(4!)^n}(-\lambda)^n\left\langle  \underbrace{x^4\cdots x^4}_n\right\rangle,$$ for that purpose we
observe that

1) $\left\langle e^{Jx}\right\rangle=e^{J^2\over 2a}$,

2) $\left\langle x^{2m+1}\right\rangle=0$  for any integer $m\ge
0$,

3) $\left\langle xx\right\rangle={\partial^2\over \partial
J^2}\left.\left\langle e^{Jx}\right\rangle\right|_{J=0}={1\over
a}$,

4) $\left\langle \underbrace{x\cdots
x}_{2m}\right\rangle={\partial^{2m}\over \partial
J^{2m}}\left.\left\langle e^{Jx}\right\rangle\right|_{J=0}$ which
is equal to the number of complete parings of $x$'s in
$\underbrace{x\cdots x}_{2m}$ times $({1\over a})^m$.

Viewing $\underbrace{x^4\cdots x^4}_{n}$ as ${\cal X}_n$ - a
collection of $n$ identical copies of $x$-cross (here $x$-cross
means a cross with each of its four legs being attached a $x$).
The topological symmetry group of ${\cal X}_n$ is
$\hbox{G}_n=(\hbox{S}_4)^n\ltimes \hbox{S}_n$ - the semi-product of $(\hbox{S}_4)^n$ with
$\hbox{S}_n$. Let $\ms{P}_n$ be the set of all possible complete parings
of $x$'s in ${\cal X}_n$. Then $\hbox{G}_n$ acts on $\ms{ P}_n$. Note
that an orbit of this action can be identified with a graph
obtained by pairing the $x$'s in ${\cal X}_n$ according to any
complete pairing in the orbit. Now, if $\Gamma$ is such a orbit or
graph (called Feynman diagram), then
\begin{eqnarray}
|\Gamma|={|\hbox{G}_n|\over |\mathrm{Aut}(\Gamma)|}
\end{eqnarray}
where $|\hbox{S}|$ denotes the number of elements in set
$\hbox{S}$ and $\mathrm{Aut}(\Gamma)$ means the subgroup of
$\hbox{G}_n$ that fixes an element in $\Gamma$, called the
symmetry group of the Feynman diagram $\Gamma$. Therefore,
$${1\over n!(4!)^n}(-\lambda)^n\left\langle  \underbrace{x^4\cdots x^4}_n\right\rangle
=\sum_{\Gamma\in\left\{
\begin{matrix}
\hbox{$4$-valent closed graphs}\cr
\hbox{with $n$-vertices}
\end{matrix}
\right\}}{1\over |\mathrm{Aut}(\Gamma)|}(-\lambda)^n({1\over
a})^{2n},$$ where ${1\over |Aut(\Gamma)|}(-\lambda)^n({1\over
a})^{2n}$ is the contribution from Feynman diagram $\Gamma$
according to the following Feynman rules:

1) To each vertex of $\Gamma$ we assign $-\lambda$,

2) To each $1$-simplex of $\Gamma$ we assign ${1\over a}$ (called
the propagator),

3) Multiply the contributions from all vertices and all
$1$-simplexes and then divided by the order of the symmetry group
of $\Gamma$.

In summary, we have
$$Y(\lambda)\sim \sum_{\Gamma\in\left\{\mbox{$4$-valent closed graphs}\right\}}
{1\over |\mathrm{Aut}(\Gamma)|}(-\lambda)^{v_\Gamma}({1\over
a})^{e_\Gamma},$$ where $v_\Gamma$ and $e_\Gamma$ are the number
of vertices and $1$-simplexes of $\Gamma$. Note that the
contribution from the empty graph is set to be $1$ by convention.
And it is tautological that
\begin{eqnarray}
\log Y(\lambda) \sim \sum_{\Gamma\in\left\{
\begin{matrix} \mbox{connected nonempty}\\
\mbox{$4$-valent closed graphs}
\end{matrix}
 \right\}}{1\over
|\mathrm{Aut}(\Gamma)|}(-\lambda)^{v_\Gamma}({1\over
a})^{e_\Gamma}.
\end{eqnarray}

It is not hard to see how to generalize all the above discussion
to the general case when other types of vertices (such as $3$-valent,
$5$-valent, \ldots) may also appear.

\bibliographystyle{unsrt}

\begin{thebibliography}{10}

\bibitem{Sma98}
S.~Smale.
\newblock Mathematical problems for the next century.
\newblock {\em Math. Intelligencer}, 20(2):7--15, 1998.

\bibitem{BCW82}
H.~Bass, E.~Connell, and D.~Wright.
\newblock The jacobian conjecture: Reduction of degree and formal expansion of
  the inverse.
\newblock {\em Bulletin of the American Mathematical Society}, 7:287--330,
  1982.

\bibitem{Kel39}
O.~H. Keller.
\newblock Ganze cremona-transformationen.
\newblock {\em Monats. Math. Physik}, 47:299--306, 1939.

\bibitem{Eng55}
W.~Engle.
\newblock Ein satz \"{u}ber ganze cremona transfirmationen der ebens.
\newblock {\em Math. Ann.}, 130:11--19, 1955.

\bibitem{Seg56-57}
B.~Segre.
\newblock Corrispondenze di mobius e transformazioni cremoniane intere.
\newblock {\em Atti della Accademia delle Scienze di Torino, Classe di Scienzo
  Fisiche, Mat. e Natural}, 91:3--19, 1956-1957.

\bibitem{Seg56}
B.~Segre.
\newblock {\em Forme differenziali e loro integrali}.
\newblock vol. II, Docet, Roma, 1956.

\bibitem{Seg60}
B.~Segre.
\newblock Variazione continua ed omotopia in geometria algebrica.
\newblock {\em Ann. Math. Pura Appl.}, 100:149--186, 1960.

\bibitem{Ess01}
A.~R.~P. van~den Essen.
\newblock {\em Progress in Mathematics, 190}.
\newblock Birkh\"{a}user Verlag, Basel, 2000. xviii+329 pp.

\bibitem{Abd02a}
A.~Abdesselam.
\newblock The jacobian conjecture as a problem of perturbative quantum field
  theory.
\newblock {\em Annales Henri Poincare}, 4:199--215, 2003.

\bibitem{Abd02b}
A.~Abdesselam.
\newblock Feynman diagrams in algebraic combinatorics.
\newblock {\em preprint}, 2002.

\bibitem{JC4SymmetricJ}
Arno van~den Essen and Sherwood Washburn.
\newblock The {Jacobian} conjecture for symmetric {J}acobian matrices.
\newblock {\em J. Pure Appl, Algebra}, 189(1--3):123--133, 2004.

\bibitem{NilpotentSymmetric}
Michiel de~Bondt and Arno van~den Essen.
\newblock Nilpotent symmetric {J}acobian matrices and the {J}acobian
  conjecture.
\newblock {\em J. Pure Appl. Algebra}, 193(1--3):61--70, 2004.

\bibitem{SymmetricReduction}
Michiel de~Bondt and Arno van~den Essen.
\newblock A reduction of the {J}acobian conjecture to the symmetric case.
\newblock Report 0308, Dept. of Math., Univ. of Nijmegen, June 2003; to appear
  in Proc. AMS.

\bibitem{SymmetricJacobian}
Franki Dillen.
\newblock Polynomials with constant {H}essian determinant.
\newblock {\em J. Pure Appl. Algebra}, 71:13--18, 1991.

\bibitem{Abh74}
S.~S. Abhyankar.
\newblock {\em Lectures in algebraic geometry}.
\newblock Purdue Univ., 1974.

\bibitem{BH86}
N.~Bleistein and R.~A. Handelsman.
\newblock {\em Asymtotic Expansions of Integrals}.
\newblock Dover Publications, Inc., New York, 1986.

\bibitem{BIZ80}
D.~Bessis, C.~Itzykson, and J.~B. Zuber.
\newblock {\em Advances in Applied Mathematics}, 1:109--157, 1980.

\end{thebibliography}

\end{document}